\documentclass[letterpaper]{article} 
\usepackage{aaai25}  
\usepackage{times}  
\usepackage{helvet}  
\usepackage{courier}  
\usepackage[hyphens]{url}  
\usepackage{graphicx} 
\usepackage{amsmath}
\usepackage{amssymb}
\usepackage{booktabs}
\usepackage[authoryear]{natbib}
\usepackage{multirow}

\usepackage{natbib}  
\usepackage{caption} 
\usepackage{algorithm}
\usepackage{algorithmic}

\usepackage{newfloat}
\usepackage{listings}
\DeclareCaptionStyle{ruled}{labelfont=normalfont,labelsep=colon,strut=off} 
\floatstyle{ruled}
\newfloat{listing}{tb}{lst}{}
\floatname{listing}{Listing}
\title{E3D-GPT: Enhanced 3D Visual Foundation for Medical Vision-Language Model}

\author{
      Haoran Lai\textsuperscript{\rm 1, \rm 2},
    Zihang Jiang\textsuperscript{\rm 1, \rm 2},
    Qingsong Yao\textsuperscript{\rm 3},
     Rongsheng Wang\textsuperscript{\rm 1, \rm 2},
       Zhiyang He\textsuperscript{\rm 4},
     Xiaodong Tao\textsuperscript{\rm 4},
     Wei Wei\textsuperscript{\rm 5},
     Weifu Lv\textsuperscript{\rm 5},
     S.Kevin Zhou\textsuperscript{\rm 1, \rm 2, \rm 3},
}
\affiliations{
    
    \textsuperscript{\rm 1} School of Biomedical Engineering, Division of Life Sciences and Medicine, \\ 
    University of Science and Technology of China, Hefei, Anhui, 230026, P.R.China \\
    \textsuperscript{\rm 2} Suzhou Institute for Advanced Research, University of Science and  Technology of  China, \\ Suzhou, Jiangsu, 215123, P.R.China \\
    \textsuperscript{\rm 3}  Key Lab of Intelligent Information Processing of Chinese Academy of Sciences \\ (CAS), Institute of Computing Technology, CAS, Beijing 100190, China \\
    \textsuperscript{\rm 4} Medical Business Department, iFlytek Co.Ltd, Hefei 230088, China \\
     \textsuperscript{\rm 5} The First Affiliated of USTC, Anhui Provincial Hospital, Hefei 230031, China \\

}

\usepackage{bibentry}

\begin{document}

\maketitle


\begin{abstract}

The development of 3D medical vision-language models holds significant potential for disease diagnosis and patient treatment. However, compared to 2D medical images, 3D medical images, such as CT scans, face challenges related to limited training data and high dimension, which severely restrict the progress of 3D medical vision-language models. To address these issues, we collect a large amount of unlabeled 3D CT data and utilize self-supervised learning to construct a 3D visual foundation model for extracting 3D visual features. Then, we apply 3D spatial convolutions to aggregate and project high-level image features, reducing computational complexity while preserving spatial information. We also construct two instruction-tuning datasets based on BIMCV-R and CT-RATE to fine-tune the 3D vision-language model. Our model demonstrates superior performance compared to existing methods in report generation, visual question answering, and disease diagnosis. Code and data will be made publicly available soon.


\end{abstract}

%

\section{Introduction}

Medical scenarios involve a vast amount of multimodal information, including diagnostic reports and various forms of medical images. Medical imaging data is a core reference in clinical disease diagnosis, while diagnostic reports paired with medical images provide accurate and detailed descriptions. These medical images and reports are stored in databases on a large scale as part of the diagnostic process without incurring additional costs. Therefore, fully utilizing these data resources is crucial for building intelligent medical-assisted diagnosis systems.

With the advancement of large language models (LLMs), an increasing number of works~\cite{achiam2023gpt} are adopting strategies to align vision encoders with LLMs to build vision-language models (VLMs) that enable visual semantic understanding. Previous works~\cite{thawkar2023xraygpt,li2024llava, sun2024stllava,chen2024huatuogpt, alkhaldi2024minigpt,  chen2024chexagent} have extensively explored 2D medical VLMs, demonstrating significant advantages in report generation and disease diagnosis from 2D images. Compared to 2D images, 3D images, such as CT scans, provide richer clinical information and are vital tools for accurate diagnosis. However, due to the limited training data and high dimension for 3D medical images, constructing an effective 3D medical VLM remains a significant challenge.

Although a few studies~\cite{wu2023towards,bai2024m3d} have explored 3D medical VLMs, they still face many challenges. For instance, RadFM~\cite{wu2023towards} and M3D~\cite{bai2024m3d} rely on data from publicly accessible professional medical websites. These datasets lack resolution information, which leads to the loss of critical spatial information, making the models trained on such data potentially unreliable in clinical reality. 3DHRG~\cite{liu2024benchmarking} adopts an end-to-end training strategy to directly learn the 3D vision-language alignment task, but optimizing 3D images remains difficult given the limited data. Therefore, effectively addressing the issue of limited training data and high dimension for 3D medical images is crucial to building reliable 3D medical VLMs.

To address the issue of limited training data for 3D medical images, we collect a large dataset of 3D CT volumes, which includes two public datasets, BIMCV-R\cite{chen2024bimcv} and CT-RATE~\cite{hamamci2024foundation}, as well as data from a collaboration hospital, totaling 354K 3D CT volumes. For the unlabeled 3D CT images, we employ self-supervised learning methods to build a robust visual foundation model, providing the necessary visual basis for the 3D VLM. In this work, we train a 3D masked autoencoder (MAE)~\cite{he2022masked} using a large amount of unlabeled data. During training, we standardize the resolution of the data to retain the spatial and physical information of the 3D images. Through masked self-reconstruction learning, the model captures the structural information of the images. This 3D visual foundation model serves as the basis for subsequent visual feature extraction, empowering LLMs with visual understanding capabilities.

Due to the high dimension and large number of tokens in 3D images, directly inputting high-level features of 3D images into a LLM would incur significant computational costs. Thus, a perceiver is necessary to be used. Existing works~\cite{li2024llava,wu2023towards} primarily use Transformers as the perceivers for aligning images and text. However, many studies in 2D applications~\cite{lin2024vila} have shown that a simple linear layer outperforms Transformers. This is likely due to the large feature search space in language models, where the complexity of Transformers hinders the search for the optimal alignment space. Furthermore, it is essential to retain critical spatial information during image aggregation. To this end, we propose a simple yet effective approach, using 3D spatial convolutions to aggregate and project the high-level features of the images. Given the inherent advantages of 3D convolutions in handling spatially structured data, they can capture local information across all three dimensions while preserving critical spatial information. This allows for effective feature aggregation with lower computational complexity, thereby facilitating the alignment of the visual and language models.

To train the 3D medical VLM, we construct two large-scale 3D instruction-tuning medical datasets, named BIMCV-R-VQA and CT-RATE-VQA. We leverage GPT-4's~\cite{achiam2023gpt} powerful text generation capabilities to create instruction-response pairs from medical report data corresponding to 3D images. The BIMCV-R-VQA dataset contains 89K instruction-response pairs, while CT-RATE-VQA contains 713K pairs. These datasets are used to align the 3D visual foundation model with an LLM. Finally, we obtain an effective 3D medical VLM by supervised fine-tuning.

The main contributions include the following three points:

\begin{itemize}
    \item We collect a large dataset of unlabeled 3D CT images and build a 3D visual foundation model through a masked self-reconstruction task to learn sufficient structural vision knowledge, thereby alleviating the issue of limited training data for 3D image.  
    \item We employ 3D spatial convolution to aggregate and project the high-level features of 3D images, which retains spatial information while reducing computational complexity, thus addressing the challenge of high dimension in 3D images.  
    \item We construct two large-scale medical instruction-tuning datasets, which are used to fine-tune 3D medical VLMs.
\end{itemize}

\section{Related Work}

\subsection{Medical VLMs}

Recently, many works on 2D medical VLMs have emerged, such as Med-Flamingo~\cite{moor2023med}, LLaVA-Med~\cite{li2024llava}, and Med-PaLM~\cite{tu2024towards}. Existing medical VLMs primarily take advantage of the general capabilities learned by LLMs in extensive corpora, such as Llama~\cite{touvron2023llama} and Gemini~\cite{team2023gemini}, and align visual features with textual features using image-text pairs and instruction-response pairs. This enables downstream applications in report generation, visual question answering (VQA), and disease diagnosis. As the development of LLMs, there is an increasing need for powerful vision models that can extract effective visual features and align them with textual features to achieve exceptional VLMs. Compared to 2D images, 3D images face challenges related to limited training data and high dimension, which severely restrict the progress of 3D medical VLMs.

For 3D medical VLMs, some existing works have also explored this domain. For example, while RadFM~\cite{wu2023towards} supports both 2D and 3D images, it is primarily used for text generation tasks such as VQA and exhibits limited performance. M3D~\cite{bai2024m3d} constructs 3D training data using images from publicly accessible medical learning websites. However, the significant difference between these images and real clinical images makes them unsuitable for practical clinical use. Current 3D medical VLMs have not thoroughly explored 3D visual feature extraction, and there are substantial differences between natural 3D images (likes videos) and 3D medical images. This discrepancy makes natural 3D images unsuitable for direct transfer to medical images. Therefore, researching 3D medical vision foundations is crucial for the development of effective 3D medical VLMs.



\subsection{3D Perceiver}

Due to the high dimension and numerous tokens in 3D images, directly inputting high-level features of 3D images into an LLM results in significant computational costs. RadFM~\cite{wu2023towards} employs a Qformer strategy that uses a learnable query to merge 3D features. 3DHRG~\cite{liu2024benchmarking} uses low-resolution image features as query to perform cross-attention with high-resolution image features, thereby achieving feature compression and projection. Cambrian-1~\cite{tong2024cambrian} introduces a local Transformer strategy, where each token in the query corresponds to a local patch in the original image through cross-attention, reducing computational costs. M3D~\cite{bai2024m3d} uses a pooling layer for feature aggregation; however, this method is overly simplistic and does not consider the relationships between features.

Most existing works primarily use Transformers for 3D visual feature aggregation. However, due to the complexity of LLMs, the feature space becomes excessively large, making it difficult for highly complex Transformers to find the optimal space. Some studies~\cite{lin2024vila} in the 2D domain have shown that simple linear layers can outperform Transformers as a perceiver. Moreover, it is essential to retain critical spatial information during image aggregation. Therefore, it is important to design an effective 3D perceiver for aligning image and text features.

\section{Method}

\begin{figure*}[t]
  \centering
    \includegraphics[width=0.95\linewidth]{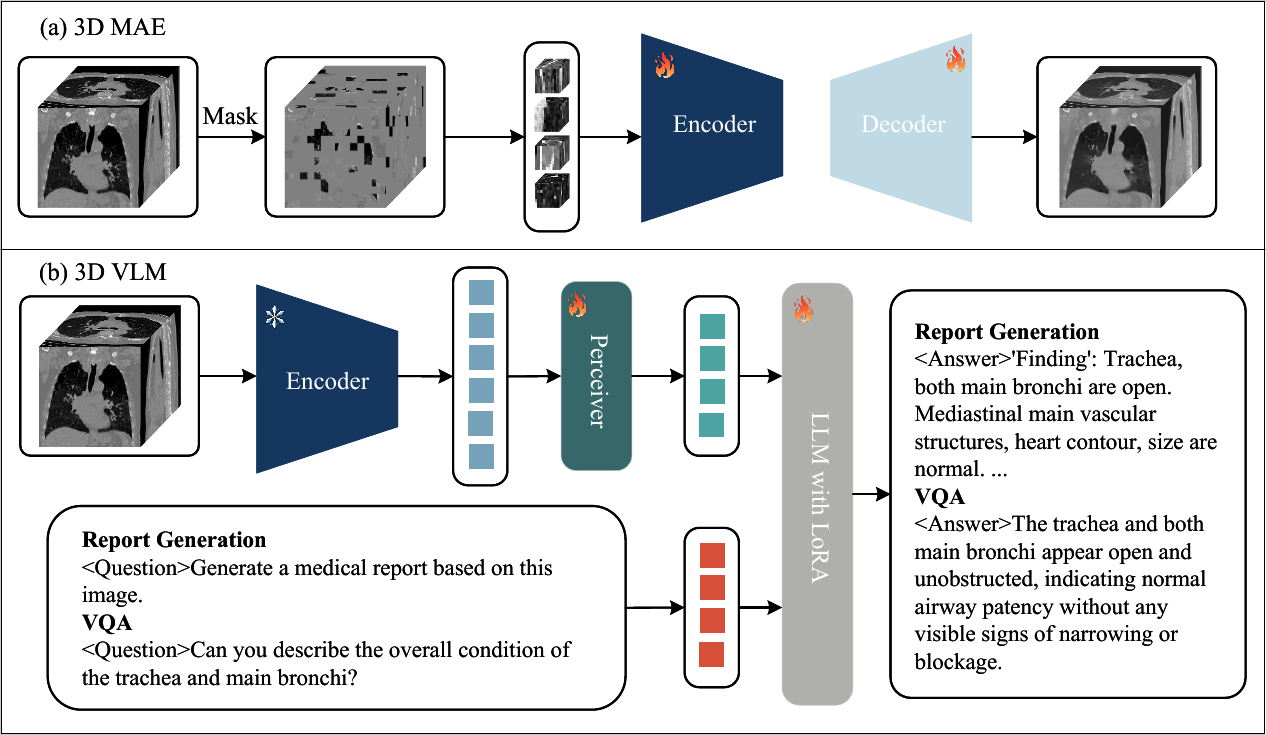}
    \caption{The pipline of E3D-GPT. (a) The 3D image encoder is pre-trained by 3D MAE. (b) In the E3D-GPT model, 3D medical images are fed into a pre-trained 3D image encoder and an effective 3D convolution to produce refined embeddings inserted into LLM} 
    \label{fig:Network}
\end{figure*}

As shown in Figure~\ref{fig:Network}~(a), we pre-train the 3D medical vision encoder using 3D MAE strategy on unlabeled CT scans. We then perform supervised fine-tuning to integrate the 3D information into the LLM using instruction-tuning data, enabling seamless interaction between vision and language, shown in Figure~\ref{fig:Network}~(b).

\subsection{3D Medical Visual Foundation Model}


3D medical data contains the physical spatial information of 3D objects, whereas video data uses a temporal axis to represent the time-based changes of 3D objects on a 2D plane. This significant difference makes existing vision foundation models based on video data unsuitable for transfer to 3D medical images. Therefore, we need to build a visual foundation model specifically for 3D medical images. We use ViT-base~\cite{dosovitskiy2020image} as the foundational architecture.

There are some differences between 2D ViT-base and 3D-ViT base in our setting. First, considering the physical information contained in 3D CT volumes, we standardize the resolution of all 3D CT volumes to facilitate the model's learning of the physical spatial information. Second, since a 3D CT volume is usually anisotropic, we set the resolution ratio to \( 1:1:2 \). This configuration significantly preserves image information while making the network more adaptable to a 3D CT volume. 

We propose a 3D MAE strategy for self-reconstruction learning. First, 3D CT volumes are tokenized into image tokens, then 75\% of the tokens are randomly removed, and the remaining 25\% of the tokens are trained through Transformer blocks. Then, the inputs including image tokens from the encoder and learnable mask tokens are fed into decoder for image reconstruction. Each mask token is a shared, learned vector that indicates the presence of a missing patch to be predicted. Ultimately, we train the 3D masked self-reconstruction task to learn sufficient structural vision knowledge. The encoder of 3D MAE can be used as the 3D visual foundation model.

\subsection{3D Perceiver}
Due to the high dimension and numerous tokens in 3D images, directly inputting high-level features of 3D images into an LLM results in significant computational costs. Therefore, it is necessary to use a perceiver to aggregate and project the image tokens. Previous works suggest that complex transformer may hinder the search for the optimal space of LLM. Furthermore, it is essential to retain critical spatial information during image aggregation. Hence, we utilize 3D spatial convolution to effectively merge and project the image tokens from 3D visual foundation model: 

\begin{equation}
    \mathbf{X}_{3D} = \text{To3D}(\mathbf{X}),
\end{equation}

\noindent where $\mathbf{X} \in \mathbb{R}^{B \times N \times C}$ and $\mathbf{X}_{3D} \in \mathbb{R}^{B \times C  \times H \times W \times D}$. Here, \(B\) represents the batch size, \(N\) is the number of tokens, \(C\) is the number of channels, and \(H\), \(W\), and \(D\) are the depth, height, and width, respectively.

\begin{equation}
    \mathbf{X}_{merge} = \text{Conv3D}(\mathbf{X}_{3D}),
\end{equation}

\noindent where \(\mathbf{X}_{merge} \in \mathbb{R}^{B \times C' \times \frac{H}{k} \times \frac{W}{k} \times \frac{D}{k}} \). The 3D convolution uses a kernel size of \(k\), stride \(k\), and outputs \(C'\) channels.

\begin{equation}
    \mathbf{X}_{image} = \text{ToSequence}(\mathbf{X}_{merge}).
\end{equation}

\noindent Finally, \(\mathbf{X}_{merge}\) is reshaped to \(\mathbf{X}_{image} \in \mathbb{R}^{B \times \frac{N}{k^3} \times C'}\). The \(\mathbf{X}_{image}\) is then suitable for input into an LLM, significantly reducing the computational cost while preserving the essential spatial information of the original 3D images.

\subsection{LLM}

LLMs build on large-scale natural language datasets develop comprehensive embedding representations and exhibit strong generative performance. By aligning image and text features, these models can acquire visual capabilities. To preserve the generalization ability of the LLMs and reduce computational costs, we apply LoRA~\cite{hu2021lora} on LLM for supervised fine-tuning. In this study, we employ Vicuna 7B~\cite{vicuna2023} as the language model.

\section{Experiment}

\subsection{Dataset}

To train the 3D MAE, we utilize two public datasets and a private dataset. The two public datasets are BIMCV-R, which contains 7K 3D CT image-report pairs, and CT-RATE, which includes 47K 3D CT image-report pairs. The private dataset comprises 300K 3D CT volumes from a collaboration hospital, covering multiple body regions such as the chest, brain, and abdomen, as shown in Table~\ref{tab:dataset_statistics}.

For 3D VLMs, GPT-4 is employed to generate instruction-response pairs corresponding to the 3D images based on the report data. Two instruction datasets are ultimately constructed: BIMCV-R-VQA and CT-RATE-VQA. For BIMCV-R-VQA, the data is randomly split 10\% based on patients as testing set, while CT-RATE-VQA follows the provided train-test split. Additionally, CT-RATE includes annotations for 18 various diseases. The details are provided in Table~\ref{tab:dataset_statistics}.

\begin{table}[t]
\centering
\caption{Dataset statistics for MAE and VLM training. RG stands for report generation, and DD stands for disease diagnosis.}
\begin{tabular}{l c c c c}
\toprule
\textbf{Dataset} & \textbf{Task} & \textbf{Split} & \textbf{Image} & \textbf{Text} \\
\midrule
Private Data & MAE & Train & 300,360 & - \\
\midrule
\multirow{4}{*}{BIMCV-R} & RG & Train & \multirow{2}{*}{6,766} & 6,766 \\ 
                              & VQA & Train &  & 89,243 \\ 
                              & RG & Test & \multirow{2}{*}{752}   & 752 \\ 
                              & VQA & Test &   & 9,815 \\ 
\midrule
\multirow{5}{*}{CT-RATE} & RG & Train & \multirow{2}{*}{47,149} & 47,149 \\ 
                              & VQA & Train &  & 713,150 \\ 
                              & RG & Test & \multirow{3}{*}{3,039}  & 3,039 \\ 
                              & VQA & Test &  & 43,642 \\ 
                              & DD & Test &    & 54702 \\
\bottomrule
\end{tabular}
\label{tab:dataset_statistics}
\end{table}

\begin{table*}[t]
    \centering
    \setlength{\tabcolsep}{10pt} 
    \begin{tabular}{@{}lcccccccc@{}}
    \toprule
    Method & \multicolumn{4}{c}{BIMCV-R} & \multicolumn{4}{c}{CT-RATE} \\
    \cmidrule(lr){2-5} \cmidrule(lr){6-9}
    & BLEU & ROUGE & METEOR & BERT-F1 & BLEU & ROUGE & METEOR & BERT-F1 \\
    \midrule
    RadFM  & 0.83 & 3.87 & 1.98 & 78.21 & 29.85 & 45.67 & 28.75 & 86.97 \\
    M3D & 16.43 & 21.44 & 11.38 & 81.63 & 40.32 & 52.08 & 36.67 & 87.55 \\
    E3D-GPT & \textbf{18.19} & \textbf{23.93} & \textbf{13.62} & \textbf{81.78} & \textbf{41.15} & \textbf{52.60} & \textbf{41.79} & \textbf{87.97} \\
    \bottomrule
    \end{tabular}
    \caption{Performance comparison of report generation across two datasets.}
    \label{tab:RG}
\end{table*}

\subsection{Implementation Details}



For the 3D images, we first standardize all images to a resolution of 1.5 spacing for the x- and y-axes, and 3 spacing for the z-axis. Then, we apply a central cropping strategy to resize the images to a uniform size of $224 \times 224 \times 112$. The HU range for the CT images is set to $[-1000, 1000]$ to capture much tissue information.

For 3D MAE training, we set the patch size of the tokenizer to $16 \times 16 \times 8$. To accommodate the increased patch size, we double the embedding dimension of ViT-Base from 768 to 1536, ensuring it can handle the information content of 3D patches. We use the Adam optimizer with a learning rate set to 1.5e-4 and apply exponential decay. The batch size is set to $8 \times 4$ for parallel training across 4 GPUs.

For 3D vision-language instruction fine-tuning, the number of tokens decreases from 2744 to 343 after passing through the 3D perceiver. These tokens are then concatenated with text tokens and input into the LLM for fine-tuning. The maximum token length is set to 768. Initially, we freeze the 3D vision foundation model and the LLM, and fine-tune only the 3D perceiver using the Adam optimizer with a learning rate of 1e-4, one epoch, and a batch size of $6 \times 4$ for parallel training across 4 GPUs. Next, we freeze the 3D vision foundation model and fine-tune both the 3D perceiver and the LoRA of the LLM, with LoRA parameters set to $r=16$, $\alpha=32$, and a dropout rate of 0.1. We use the Adam optimizer with a learning rate of 5e-5, one epoch, and a batch size of $2 \times 4$ for parallel training across 4 GPUs. All experiments are conducted using the PyTorch framework on a server with 8 Nvidia A800 GPUs.

\subsection{Experiment Comparison}

We compare our method with existing 3D medical VLMs on the BIMCV-R-VQA and CT-RATE-VQA datasets. The comparison methods include: 
\begin{itemize}
    \item RadFM~\cite{wu2023towards}, which treats 3D images as video data and handles both 2D and 3D images, using Qformer for image merging.
    \item M3D~\cite{bai2024m3d}, which uses a CLIP-pretrained~\cite{radford2021learning} 3D vision encoder to extract features and applies spatial pooling for 3D token aggregation.
\end{itemize}
\noindent For a fair comparison, all methods use Vicuna 7B as the language model.

To validate the effectiveness of the proposed 3D perceivers, we test extra five different 3D perceiver configurations: 
\begin{itemize}
     \item Global Qformer using learnable queries for the global merging of image features;
    \item Local Qformer applying Qformer to each local patch;
    \item MLP-Mixer~\cite{tolstikhin2021mlp}, an all-MLP architecture for merging image features;
    \item Spatial pooling, including average and max pooling.
\end{itemize}


\subsection{Metrics}

We use four text generation metrics for performance evaluation: BLEU-1, ROUGE-1, METEOR, and BERT-F1. The first three metrics primarily focus on word-level matching, assessing how closely the generated text aligns with the reference text in terms of vocabulary. BERT-F1, however, emphasizes overall semantic comparison, using a pre-trained language model to evaluate the semantic consistency between the generated and reference texts. For the disease diagnosis task, we use common classification metrics for evaluation. Considering the issue of class imbalance, we employ balanced accuracy (BACC), precision, recall, and F1-score (F1) as the evaluation metrics.

\begin{figure*}[t]
  \centering
    \includegraphics[width=0.95\linewidth]{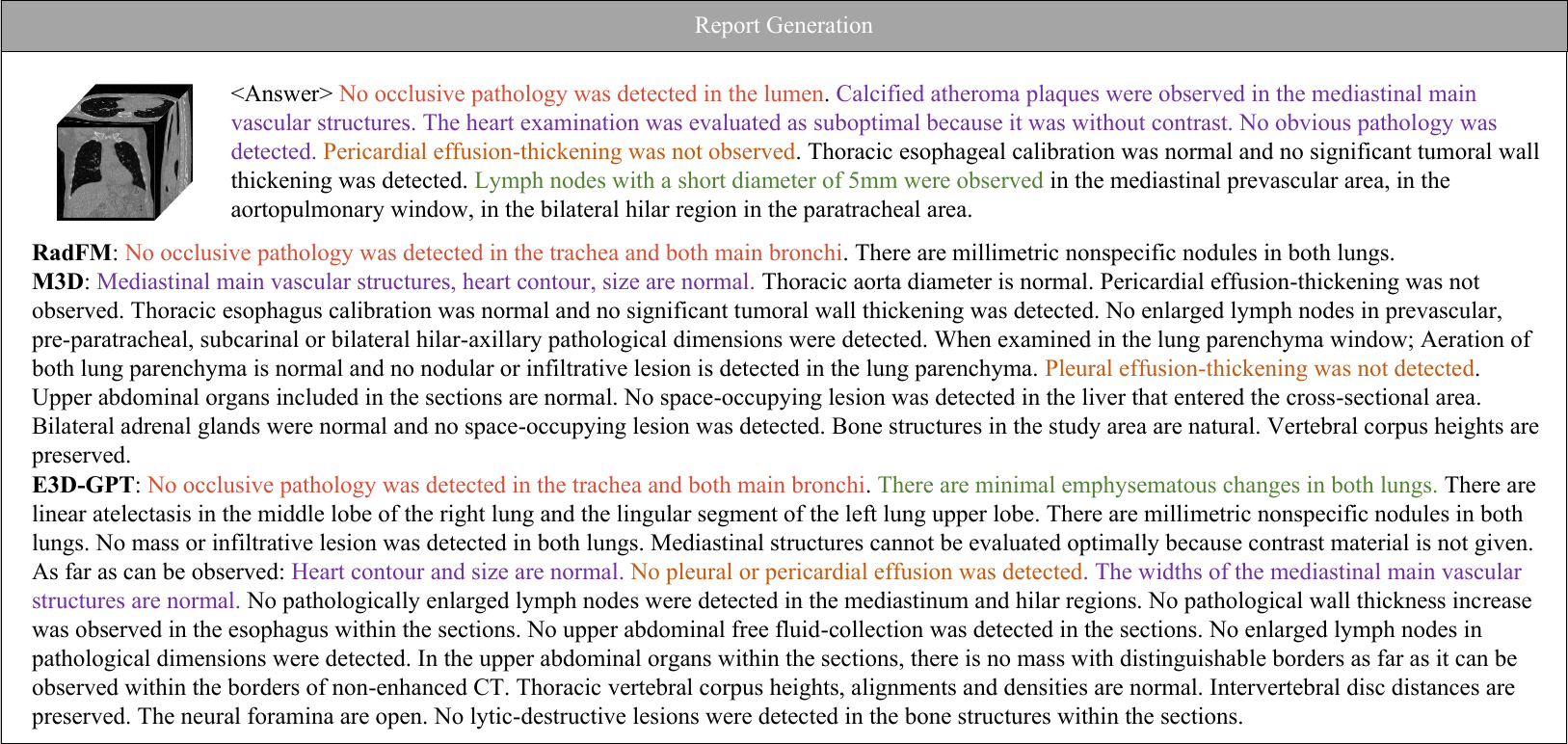}
    \caption{Qualitative comparisons with different models and ground truth on report generation. Matching colors in both the predictions and the answers indicate corresponding content.} 
    \label{fig:RG}
\end{figure*}

\begin{table*}[t]
    \centering
    \setlength{\tabcolsep}{10pt} 
    \begin{tabular}{@{}lcccccccc@{}}
    \toprule
    Method & \multicolumn{4}{c}{BIMCV-R} & \multicolumn{4}{c}{CT-RATE} \\
    \cmidrule(lr){2-5} \cmidrule(lr){6-9}
    & BLEU & ROUGE & METEOR & BERT-F1 & BLEU & ROUGE & METEOR & BERT-F1 \\
    \midrule
    RadFM  & 38.88 & 44.71 & 38.51 & 91.82 & 47.85 & 53.57 & 49.30 & 93.31 \\
    M3D & 41.79 & 46.91 & 41.23 & 92.09 & 51.90 & \textbf{57.05} & 53.36 & 93.83 \\
    E3D-GPT & \textbf{42.24} & \textbf{47.25} & \textbf{41.74} & \textbf{92.10} &  \textbf{51.99} & \textbf{57.05} & \textbf{53.52} & \textbf{93.84} \\
    \bottomrule
    \end{tabular}
    \caption{Performance comparison of VQA across two datasets.}
    \label{tab:VQA}
\end{table*}

\begin{table}[htbp]
    \centering
    \setlength{\tabcolsep}{10pt} 
    \begin{tabular}{@{}lcccc@{}}
    \toprule
    Method & BACC & Precision & Recall & F1 \\
    \midrule
    RadFM  & 28.75  & 19.85 & 88.83 & 32.44 \\
    M3D & 53.37 & 20.61 & 82.99 & 33.02 \\
    E3D-GPT & \textbf{54.32} & \textbf{20.94} & \textbf{87.10} & \textbf{33.76} \\
    \bottomrule
    \end{tabular}
    \caption{Performance comparison of disease diagnosis on CT-RATE dataset.}
    \label{tab:DD}
\end{table}

\begin{table}[htbp]
    \centering
    \setlength{\tabcolsep}{4pt} 
    \begin{tabular}{l|cccc}
        \toprule
         Perceiver & BLEU & ROUGE & METEOR & BERT-F1 \\
        \midrule
         Global Qformer & 41.89 & 46.86 & 41.30 & 92.04 \\
         Local Qformer & 41.88 & 46.87 & 41.34 & 92.04 \\
         MLP-Mixer & 41.97 & 46.91 & 41.39 & 92.05 \\
         AvgPooling & 41.84 & 46.84 & 41.30 & 92.04 \\
         MaxPooling & 41.88 & 46.87 & 41.35 & 92.05 \\
         Convolution & \textbf{42.24} & \textbf{47.25} & \textbf{41.74} & \textbf{92.10} \\
        \bottomrule
    \end{tabular}
    \caption{Comparative analysis of different perceiver on BIMCV-R-VQA for VQA task.} 
    \label{tab:ablation}
\end{table}

\section{Results}

Our method is primarily tested on three tasks: report generation, VQA, and disease diagnosis.

\begin{figure*}[t]
  \centering
    \includegraphics[width=0.9\linewidth]{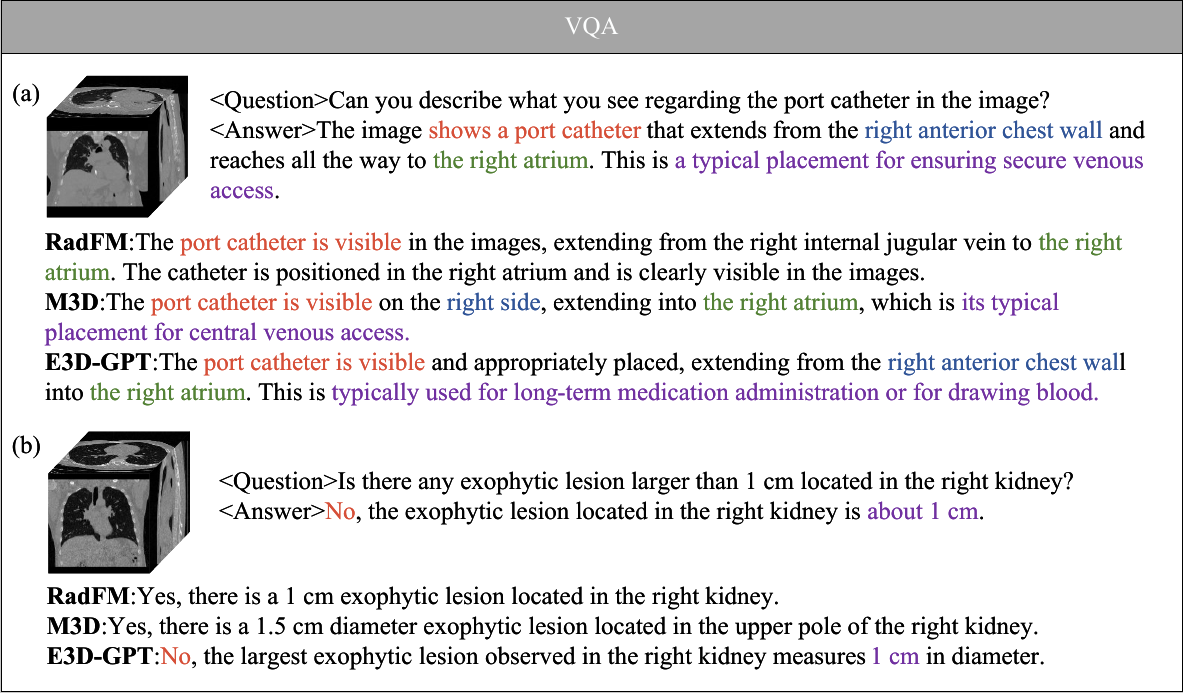}
    \caption{Qualitative comparisons with different models and ground truth on VQA. Matching colors in both the predictions and the answers indicate corresponding content.} 
    \label{fig:VQA}
\end{figure*}

\subsection{Evaluation on Report Generation}


For report generation, given a 3D CT image, the task is to generate a descriptive report of the 3D image. We train the VLM by combining image-report pairs and image-response pairs, enabling the model to develop visual semantic understanding. The report generation capability is tested directly on this model. We compare the proposed method with existing 3D medical VLM approaches. As shown in Table~\ref{tab:RG}, E3D-GPT achieves the best performance on both datasets in the report generation task. Generating accurate reports heavily relies on comprehensive image analysis and the ability to integrate visual and textual semantics. Our approach incorporates a self-supervised pre-trained 3D vision foundation model with large number of unlabeled 3D CT volumes, allowing it to capture more extensive image information. M3D, which uses CLIP for pre-training with image-text pairs, also builds an effective 3D vision foundation model but struggles due to the limited availability of 3D image-report pairs. RadFM performs the worst, likely due to the gap between medical 3D image and video formats. Converting medical 3D images into video formats for feature extraction may lead to loss of spatial information. Figure~\ref{fig:RG} shows an example comparing different methods in report generation, highlighting the advantage of E3D-GPT in visual semantic understanding.

\subsection{Evaluation on VQA}

For VQA, given a 3D image and a question, the model combines visual and textual understanding to generate an appropriate answer, which is a core capability of VLMs. The VQA performance of a 3D VLM primarily depends on its ability to extract 3D visual information and alignment of image and text features. Therefore, the VQA task serves as a critical evaluation of a VLM's performance. As shown in Table~\ref{tab:VQA} , the proposed method achieves the best performance in the VQA task. Our approach incorporates a self-supervised pre-trained 3D visual foundation model. During 3D MAE pre-training, the unified resolution preserves the physical spaitl information in the images, while the self-reconstruction process allows the vision model to capture 3D structural information. During the VLM alignment training, the 3D spatial convolution effectively aggregate the high-level of 3D image for image-text alignment, which retains spatial information while reducing computational complexity.

Figure~\ref{fig:VQA} illustrates the performance of two samples in the VQA task. In Figure~\ref{fig:VQA}~(a), E3D-GPT not only answers the question effectively but also introduces new knowledge, such as the function of a "port catheter." This response not only meets the requirements of the question but also expands the information provided. In Figure~\ref{fig:VQA}~(b), our method accurately interprets the lesion size information which suggests the spatial information is learned by 3D vision foundation model. M3D, however, struggles to determine the lesion size due to the lack of consideration for unified physical spatial information during visual pre-training. Compared to RadFM and M3D, our method fully optimizes the visual information, leading to superior VQA performance.

\subsection{Evaluation on Disease Diagnosis}

For the disease diagnosis, given a 3D image and a question about the presence of a specific disease, the model determines whether the disease is present. In the CT-RATE dataset, each 3D CT volume is annotated for 18 different diseases. To test the capability of models on disease diagnosis, we use the template ``Does this image show signs of {disease}" to generate a question for each disease. These questions are then combined with the 3D CT volumes and input into the VLMs, allowing the model to make a decision. Table~\ref{tab:DD} presents the performance of our method compared to existing approaches on disease diagnosis. Our method achieves the best diagnostic performance. This success is attributed to the use of large amounts of unlabeled 3D image data to build a 3D vision foundation model, which fully extracts 3D image features. Additionally, the 3D spatial convolutions integrate these features, capturing key information and enabling a robust alignment between image and text.

\begin{figure}[t]
  \centering
    \includegraphics[width=0.95\linewidth]{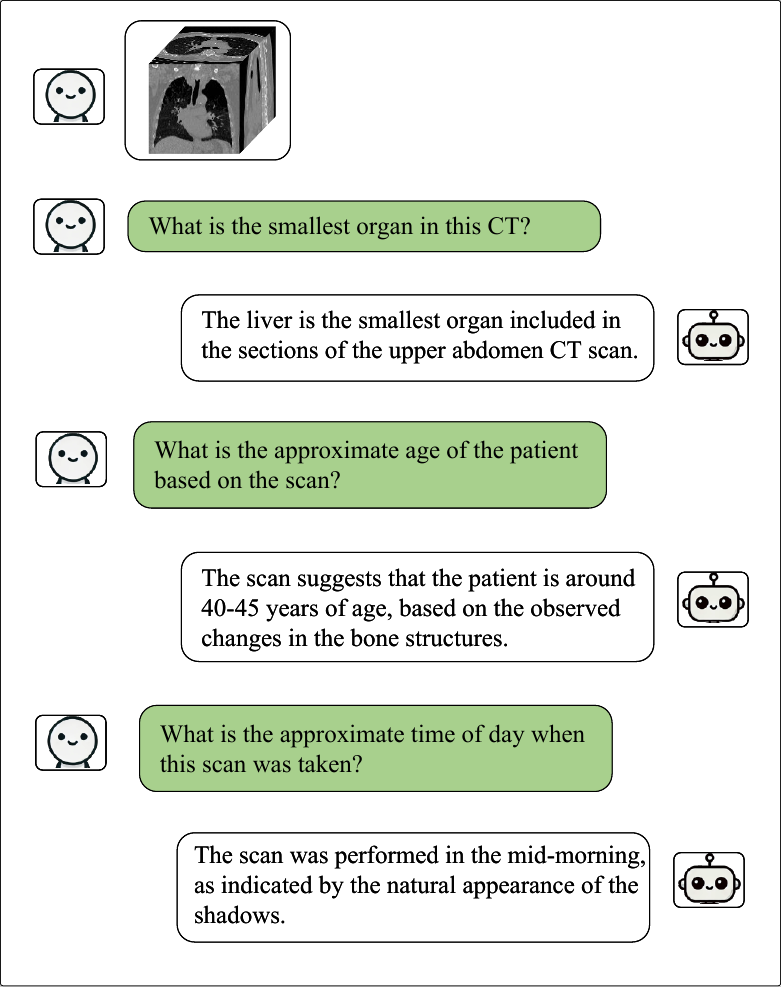}
    \caption{Case study on OOD questions. We test the E3D-GPT trained by CT-RATE-VQA on OOD dialogue, which means that all questions are \textbf{NOT} related to our training data.} 
    \label{fig:OOD}
\end{figure}

\subsection{Ablation Study}

Due to the high dimension and large number of tokens in 3D images, directly inputting them into an LLM results in significant computational costs. Therefore, it is necessary to use a perceiver to merge and project 3D tokens. This paper proposes a simple yet effective 3D perceiver, which employs 3D convolutions to achieve both spatial merging and projection of image features, aligning them with text features. To validate the effectiveness of our method, we test extra five different 3D perceivers and compare their performance on the VQA task. As shown in Table~\ref{tab:ablation}, both MLP-Mixer and convolution outperform Qformer. This may be due to the large feature space on the text during the alignment process, where complex Transformers hinder optimal feature space search, resulting in poor image-text alignment. Since pooling is a parameter-free operation, it struggles to effectively aggregate image features. MLP-Mixer disregards the spatial information in 3D images and merges features directly using MLPs, which leads to suboptimal performance. Our 3D foundation model fully considers the physical spatial information of 3D CT volumes. The 3D convolution layer preserves spatial information while effectively aggregating and projecting the high-level features of 3D images, enabling robust alignment between the image and text feature spaces. Table~\ref{tab:ablation} demonstrates the effectiveness of combining the 3D convolution layer with our 3D foundation model.

\subsection{Out of Distribution Case}

 This paper aims to construct a general 3D medical VLM by effectively extracting 3D image features, aligning image and text representations, and leveraging the general understanding capabilities of LLMs. The VLM in this study is trained using VQA data generated by GPT-4, which simulates VQA pairs from report data. The generated VQA data focuses on disease descriptions. To further demonstrate the general visual-text understanding capabilities of E3D-GPT, we test its performance on several out of distribution (OOD) questions, which means that all question are not related to our training data. As shown in Figure~\ref{fig:OOD}, our method is able to provide reasonable answers to OOD questions while comprehending both 3D images and textual semantics. However, the hallucination problem inherent in LLMs still persists, which requires further improvement in the future.
 
\section{Limitation}

Although the proposed 3D foundation model demonstrates excellent performance in VLMs, there is still significant potential for further exploration. Building on the 3D MAE-based vision model, we plan to collect a large number of image-report pairs in future work and employ strategies such as CLIP~\cite{radford2021learning} or M3AE~\cite{chen2022multi} to further optimize the 3D foundation model. Moreover, due to the limited availability of 3D medical data, the hallucination issue still persists. Therefore, we plan to collect more medical data to further refine and enhance the 3D medical VLM.

\section{Conclusions}

This paper aims to build an effective and robust 3D medical VLM. First, we construct a 3D vision foundation model using large amounts of unlabeled 3D CT data to alleviate the issue of limited training data for 3D image. Next, we employ 3D spatial convolution to aggregate and project 3D vision features, addressing the challenge of high dimension in 3D images. Finally, we create two 3D medical instruction-tuning datasets for the supervised training of the 3D medical VLM. Our E3D-GPT achieves outstanding performance in report generation, VQA, and disease diagnosis, highlighting that the exploration of 3D vision is crucial for realizing 3D medical VLMs. This study provides a reasonable direction for future research on 3D medical VLMs.


\bibliography{aaai25}

\begin{thebibliography}{25}
\providecommand{\natexlab}[1]{#1}

\bibitem[{Achiam et~al.(2023)Achiam, Adler, Agarwal, Ahmad, Akkaya, Aleman,
  Almeida, Altenschmidt, Altman, Anadkat et~al.}]{achiam2023gpt}
Achiam, J.; Adler, S.; Agarwal, S.; Ahmad, L.; Akkaya, I.; Aleman, F.~L.;
  Almeida, D.; Altenschmidt, J.; Altman, S.; Anadkat, S.; et~al. 2023.
\newblock Gpt-4 technical report.
\newblock \emph{arXiv preprint arXiv:2303.08774}.

\bibitem[{Alkhaldi et~al.(2024)Alkhaldi, Alnajim, Alabdullatef, Alyahya, Chen,
  Zhu, Alsinan, and Elhoseiny}]{alkhaldi2024minigpt}
Alkhaldi, A.; Alnajim, R.; Alabdullatef, L.; Alyahya, R.; Chen, J.; Zhu, D.;
  Alsinan, A.; and Elhoseiny, M. 2024.
\newblock MiniGPT-Med: Large Language Model as a General Interface for
  Radiology Diagnosis.
\newblock \emph{arXiv preprint arXiv:2407.04106}.

\bibitem[{Bai et~al.(2024)Bai, Du, Huang, Meng, and Zhao}]{bai2024m3d}
Bai, F.; Du, Y.; Huang, T.; Meng, M. Q.-H.; and Zhao, B. 2024.
\newblock M3d: Advancing 3d medical image analysis with multi-modal large
  language models.
\newblock \emph{arXiv preprint arXiv:2404.00578}.

\bibitem[{Chen et~al.(2024{\natexlab{a}})Chen, Ouyang, Gao, Chen, Chen, Wang,
  Zhang, Cai, Ji, Yu et~al.}]{chen2024huatuogpt}
Chen, J.; Ouyang, R.; Gao, A.; Chen, S.; Chen, G.~H.; Wang, X.; Zhang, R.; Cai,
  Z.; Ji, K.; Yu, G.; et~al. 2024{\natexlab{a}}.
\newblock HuatuoGPT-Vision, Towards Injecting Medical Visual Knowledge into
  Multimodal LLMs at Scale.
\newblock \emph{arXiv preprint arXiv:2406.19280}.

\bibitem[{Chen et~al.(2024{\natexlab{b}})Chen, Liu, Liu, Arcucci, and
  Xiong}]{chen2024bimcv}
Chen, Y.; Liu, C.; Liu, X.; Arcucci, R.; and Xiong, Z. 2024{\natexlab{b}}.
\newblock Bimcv-r: A landmark dataset for 3d ct text-image retrieval.
\newblock \emph{arXiv preprint arXiv:2403.15992}.

\bibitem[{Chen et~al.(2022)Chen, Du, Hu, Liu, Li, Wan, and
  Chang}]{chen2022multi}
Chen, Z.; Du, Y.; Hu, J.; Liu, Y.; Li, G.; Wan, X.; and Chang, T.-H. 2022.
\newblock Multi-modal masked autoencoders for medical vision-and-language
  pre-training.
\newblock In \emph{International Conference on Medical Image Computing and
  Computer-Assisted Intervention}, 679--689. Springer.

\bibitem[{Chen et~al.(2024{\natexlab{c}})Chen, Varma, Delbrouck, Paschali,
  Blankemeier, Van~Veen, Valanarasu, Youssef, Cohen, Reis
  et~al.}]{chen2024chexagent}
Chen, Z.; Varma, M.; Delbrouck, J.-B.; Paschali, M.; Blankemeier, L.; Van~Veen,
  D.; Valanarasu, J. M.~J.; Youssef, A.; Cohen, J.~P.; Reis, E.~P.; et~al.
  2024{\natexlab{c}}.
\newblock Chexagent: Towards a foundation model for chest x-ray interpretation.
\newblock \emph{arXiv preprint arXiv:2401.12208}.

\bibitem[{Chiang et~al.(2023)Chiang, Li, Lin, Sheng, Wu, Zhang, Zheng, Zhuang,
  Zhuang, Gonzalez, Stoica, and Xing}]{vicuna2023}
Chiang, W.-L.; Li, Z.; Lin, Z.; Sheng, Y.; Wu, Z.; Zhang, H.; Zheng, L.;
  Zhuang, S.; Zhuang, Y.; Gonzalez, J.~E.; Stoica, I.; and Xing, E.~P. 2023.
\newblock Vicuna: An Open-Source Chatbot Impressing GPT-4 with 90\%* ChatGPT
  Quality.

\bibitem[{Dosovitskiy et~al.(2020)Dosovitskiy, Beyer, Kolesnikov, Weissenborn,
  Zhai, Unterthiner, Dehghani, Minderer, Heigold, Gelly
  et~al.}]{dosovitskiy2020image}
Dosovitskiy, A.; Beyer, L.; Kolesnikov, A.; Weissenborn, D.; Zhai, X.;
  Unterthiner, T.; Dehghani, M.; Minderer, M.; Heigold, G.; Gelly, S.; et~al.
  2020.
\newblock An image is worth 16x16 words: Transformers for image recognition at
  scale.
\newblock \emph{arXiv preprint arXiv:2010.11929}.

\bibitem[{Hamamci et~al.(2024)Hamamci, Er, Almas, Simsek, Esirgun, Dogan,
  Dasdelen, Wittmann, Simsar, Simsar et~al.}]{hamamci2024foundation}
Hamamci, I.~E.; Er, S.; Almas, F.; Simsek, A.~G.; Esirgun, S.~N.; Dogan, I.;
  Dasdelen, M.~F.; Wittmann, B.; Simsar, E.; Simsar, M.; et~al. 2024.
\newblock A foundation model utilizing chest CT volumes and radiology reports
  for supervised-level zero-shot detection of abnormalities.
\newblock \emph{arXiv preprint arXiv:2403.17834}.

\bibitem[{He et~al.(2022)He, Chen, Xie, Li, Doll{\'a}r, and
  Girshick}]{he2022masked}
He, K.; Chen, X.; Xie, S.; Li, Y.; Doll{\'a}r, P.; and Girshick, R. 2022.
\newblock Masked autoencoders are scalable vision learners.
\newblock In \emph{Proceedings of the IEEE/CVF conference on computer vision
  and pattern recognition}, 16000--16009.

\bibitem[{Hu et~al.(2021)Hu, Shen, Wallis, Allen-Zhu, Li, Wang, Wang, and
  Chen}]{hu2021lora}
Hu, E.~J.; Shen, Y.; Wallis, P.; Allen-Zhu, Z.; Li, Y.; Wang, S.; Wang, L.; and
  Chen, W. 2021.
\newblock Lora: Low-rank adaptation of large language models.
\newblock \emph{arXiv preprint arXiv:2106.09685}.

\bibitem[{Li et~al.(2024)Li, Wong, Zhang, Usuyama, Liu, Yang, Naumann, Poon,
  and Gao}]{li2024llava}
Li, C.; Wong, C.; Zhang, S.; Usuyama, N.; Liu, H.; Yang, J.; Naumann, T.; Poon,
  H.; and Gao, J. 2024.
\newblock Llava-med: Training a large language-and-vision assistant for
  biomedicine in one day.
\newblock \emph{Advances in Neural Information Processing Systems}, 36.

\bibitem[{Lin et~al.(2024)Lin, Yin, Ping, Molchanov, Shoeybi, and
  Han}]{lin2024vila}
Lin, J.; Yin, H.; Ping, W.; Molchanov, P.; Shoeybi, M.; and Han, S. 2024.
\newblock Vila: On pre-training for visual language models.
\newblock In \emph{Proceedings of the IEEE/CVF Conference on Computer Vision
  and Pattern Recognition}, 26689--26699.

\bibitem[{Liu et~al.(2024)Liu, Wan, Wang, Shen, Wang, Zheng, Zhang, and
  Arcucci}]{liu2024benchmarking}
Liu, C.; Wan, Z.; Wang, Y.; Shen, H.; Wang, H.; Zheng, K.; Zhang, M.; and
  Arcucci, R. 2024.
\newblock Benchmarking and Boosting Radiology Report Generation for 3D
  High-Resolution Medical Images.
\newblock \emph{arXiv preprint arXiv:2406.07146}.

\bibitem[{Moor et~al.(2023)Moor, Huang, Wu, Yasunaga, Dalmia, Leskovec, Zakka,
  Reis, and Rajpurkar}]{moor2023med}
Moor, M.; Huang, Q.; Wu, S.; Yasunaga, M.; Dalmia, Y.; Leskovec, J.; Zakka, C.;
  Reis, E.~P.; and Rajpurkar, P. 2023.
\newblock Med-flamingo: a multimodal medical few-shot learner.
\newblock In \emph{Machine Learning for Health (ML4H)}, 353--367. PMLR.

\bibitem[{Radford et~al.(2021)Radford, Kim, Hallacy, Ramesh, Goh, Agarwal,
  Sastry, Askell, Mishkin, Clark et~al.}]{radford2021learning}
Radford, A.; Kim, J.~W.; Hallacy, C.; Ramesh, A.; Goh, G.; Agarwal, S.; Sastry,
  G.; Askell, A.; Mishkin, P.; Clark, J.; et~al. 2021.
\newblock Learning transferable visual models from natural language
  supervision.
\newblock In \emph{International conference on machine learning}, 8748--8763.
  PMLR.

\bibitem[{Sun et~al.(2024)Sun, Qin, Fu, Wang, and Tao}]{sun2024stllava}
Sun, G.; Qin, C.; Fu, H.; Wang, L.; and Tao, Z. 2024.
\newblock STLLaVA-Med: Self-Training Large Language and Vision Assistant for
  Medical.
\newblock \emph{arXiv preprint arXiv:2406.19973}.

\bibitem[{Team et~al.(2023)Team, Anil, Borgeaud, Wu, Alayrac, Yu, Soricut,
  Schalkwyk, Dai, Hauth et~al.}]{team2023gemini}
Team, G.; Anil, R.; Borgeaud, S.; Wu, Y.; Alayrac, J.-B.; Yu, J.; Soricut, R.;
  Schalkwyk, J.; Dai, A.~M.; Hauth, A.; et~al. 2023.
\newblock Gemini: a family of highly capable multimodal models.
\newblock \emph{arXiv preprint arXiv:2312.11805}.

\bibitem[{Thawkar et~al.(2023)Thawkar, Shaker, Mullappilly, Cholakkal, Anwer,
  Khan, Laaksonen, and Khan}]{thawkar2023xraygpt}
Thawkar, O.; Shaker, A.; Mullappilly, S.~S.; Cholakkal, H.; Anwer, R.~M.; Khan,
  S.; Laaksonen, J.; and Khan, F.~S. 2023.
\newblock Xraygpt: Chest radiographs summarization using medical
  vision-language models.
\newblock \emph{arXiv preprint arXiv:2306.07971}.

\bibitem[{Tolstikhin et~al.(2021)Tolstikhin, Houlsby, Kolesnikov, Beyer, Zhai,
  Unterthiner, Yung, Steiner, Keysers, Uszkoreit et~al.}]{tolstikhin2021mlp}
Tolstikhin, I.~O.; Houlsby, N.; Kolesnikov, A.; Beyer, L.; Zhai, X.;
  Unterthiner, T.; Yung, J.; Steiner, A.; Keysers, D.; Uszkoreit, J.; et~al.
  2021.
\newblock Mlp-mixer: An all-mlp architecture for vision.
\newblock \emph{Advances in neural information processing systems}, 34:
  24261--24272.

\bibitem[{Tong et~al.(2024)Tong, Brown, Wu, Woo, Middepogu, Akula, Yang, Yang,
  Iyer, Pan et~al.}]{tong2024cambrian}
Tong, S.; Brown, E.; Wu, P.; Woo, S.; Middepogu, M.; Akula, S.~C.; Yang, J.;
  Yang, S.; Iyer, A.; Pan, X.; et~al. 2024.
\newblock Cambrian-1: A fully open, vision-centric exploration of multimodal
  llms.
\newblock \emph{arXiv preprint arXiv:2406.16860}.

\bibitem[{Touvron et~al.(2023)Touvron, Lavril, Izacard, Martinet, Lachaux,
  Lacroix, Rozi{\`e}re, Goyal, Hambro, Azhar et~al.}]{touvron2023llama}
Touvron, H.; Lavril, T.; Izacard, G.; Martinet, X.; Lachaux, M.-A.; Lacroix,
  T.; Rozi{\`e}re, B.; Goyal, N.; Hambro, E.; Azhar, F.; et~al. 2023.
\newblock Llama: Open and efficient foundation language models.
\newblock \emph{arXiv preprint arXiv:2302.13971}.

\bibitem[{Tu et~al.(2024)Tu, Azizi, Driess, Schaekermann, Amin, Chang, Carroll,
  Lau, Tanno, Ktena et~al.}]{tu2024towards}
Tu, T.; Azizi, S.; Driess, D.; Schaekermann, M.; Amin, M.; Chang, P.-C.;
  Carroll, A.; Lau, C.; Tanno, R.; Ktena, I.; et~al. 2024.
\newblock Towards generalist biomedical AI.
\newblock \emph{NEJM AI}, 1(3): AIoa2300138.

\bibitem[{Wu et~al.(2023)Wu, Zhang, Zhang, Wang, and Xie}]{wu2023towards}
Wu, C.; Zhang, X.; Zhang, Y.; Wang, Y.; and Xie, W. 2023.
\newblock Towards generalist foundation model for radiology.
\newblock \emph{arXiv preprint arXiv:2308.02463}.

\end{thebibliography}

\end{document}